\documentclass[useAMS]{mn2e}

\usepackage{latexsym,graphicx}

\usepackage{color}

\newcommand\kms{{\rm\,km\,s^{-1}}}
\newcommand\msun{\rm\,M_\odot}

\newcommand\hii{H\,{\sc ii} \,}

\def\apgt{\ {\raise-.5ex\hbox{$\buildrel>\over\sim$}}\ }
\def\aplt{\ {\raise-.5ex\hbox{$\buildrel<\over\sim$}}\ }

\title[Two massive stars ejected from NGC\,3603]{Two massive stars possibly ejected from NGC\,3603 via a three-body encounter}
\author[V.V.Gvaramadze et al.]
{V. V.~Gvaramadze,$^{1,2}$\thanks{E-mail: vgvaram@mx.iki.rssi.ru
(VVG); akniazev@saao.ac.za (AYK); andrenicolas.chene@gmail.com
(A-NC); oschnurr@aip.de (OS)}
         A. Y.~Kniazev,$^{3,4,1}$ A.-N.~Chen\'{e},$^{5,6}$ and O.~Schnurr$^{7}$\\
        $^{1}$Sternberg Astronomical Institute, Lomonosov Moscow State University, Universitetskij Pr. 13, Moscow 119992, Russia\\
        $^{2}$Isaac Newton Institute of Chile, Moscow Branch, Universitetskij Pr. 13, Moscow 119992,
        Russia \\
        $^{3}$South African Astronomical Observatory, PO Box 9, 7935 Observatory, Cape Town,
        South Africa \\
        $^{4}$Southern African Large Telescope Foundation, PO Box 9, 7935 Observatory, Cape Town,
        South Africa \\
        $^{5}$Departamento de F\'{i}sica y Astronom\'{i}a, Universidad de Valpara\'{i}so, Av. Gran
        Breta\~{n}a 1111, Playa Ancha, Casilla 5030, Chile \\
        $^{6}$Departamento de Astronom\'{i}a, Universidad de Concepci\'{o}n, Casilla 160-C,
        Chile\\
        $^{7}$Astrophysikalisches Institut Potsdam, An der Sternwarte 16, 14482 Potsdam, Germany \\
               }
\begin{document}

\date{Accepted 2012 November 26.  Received 2012 November 25; in original form 2012 October 30}

\maketitle

\label{firstpage}

\begin{abstract}

We report the discovery of a bow-shock-producing star in the
vicinity of the young massive star cluster NGC\,3603 using
archival data of the {\it Spitzer Space Telescope}. Follow-up
optical spectroscopy of this star with Gemini-South led to its
classification as O6\,V. The orientation of the bow shock and the
distance to the star (based on its spectral type) suggest that the
star was expelled from the cluster, while the young age of the
cluster ($\sim$2 Myr) implies that the ejection was caused by a
dynamical few-body encounter in the cluster's core. The relative
position on the sky of the O6\,V star and a recently discovered
O2\,If*/WN6 star (located on the opposite side of NGC\,3603)
allows us to propose that both objects were ejected from the
cluster via the same dynamical event -- a three-body encounter
between a single (O6\,V) star and a massive binary (now the
O2\,If*/WN6 star). If our proposal is correct, then one can
``weigh" the O2\,If*/WN6 star using the conservation of the linear
momentum. Given a mass of the O6\,V star of $\approx$$30 \,
\msun$, we found that at the moment of ejection the mass of the
O2\,If*/WN6 star was $\approx$$175 \, \msun$. Moreover, the
observed X-ray luminosity of the O2\,If*/WN6 star (typical of a
single star) suggests that the components of this originally
binary system have merged (e.g., because of encounter hardening).

\end{abstract}


\section{Introduction}
\label{sec:int}

\begin{figure*}
\begin{minipage}[h]{0.435\linewidth}
\center{\includegraphics[width=1\linewidth]{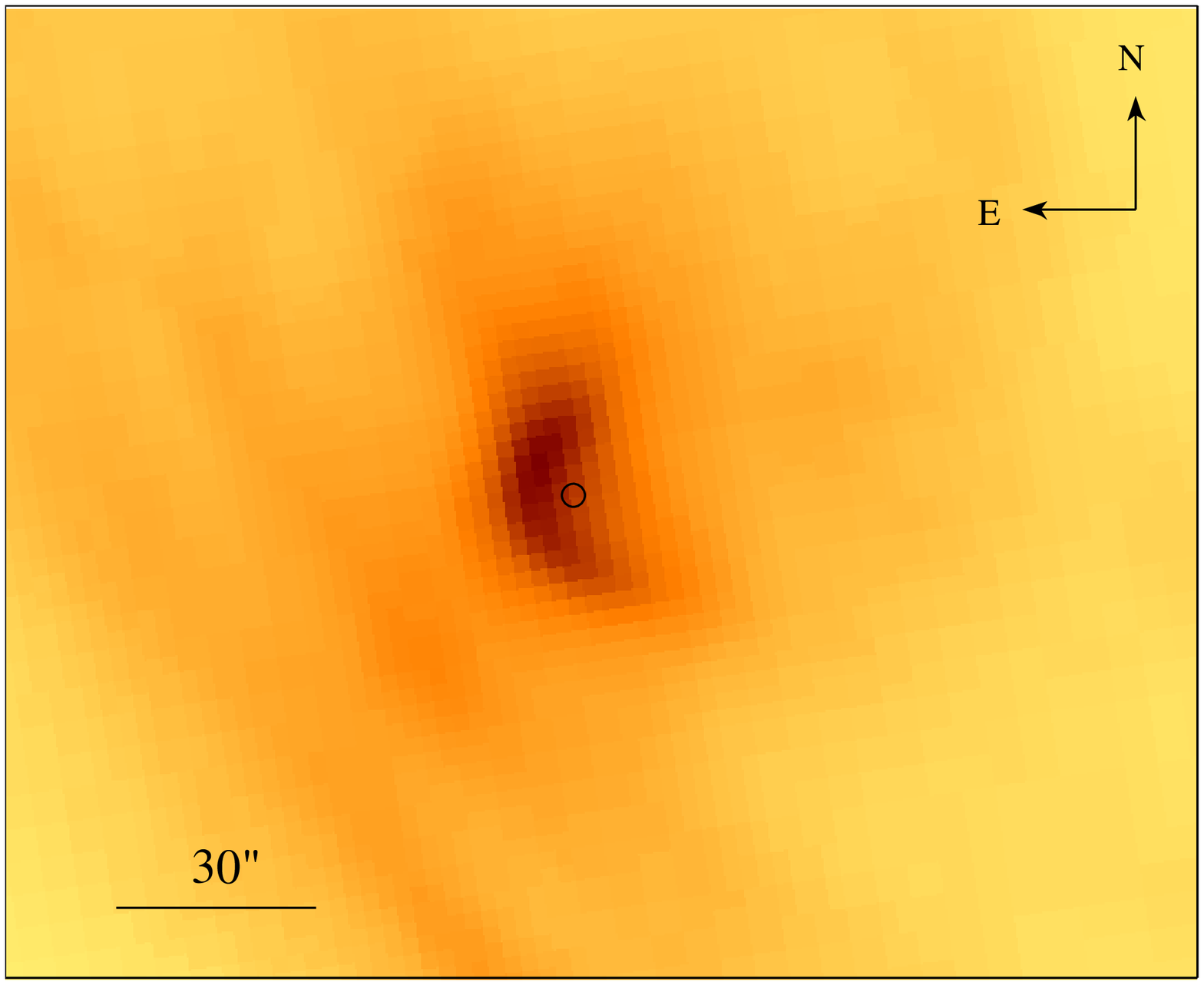}} \\
\end{minipage}
\hfill
\begin{minipage}[h]{0.54\linewidth}
\center{\includegraphics[width=1\linewidth]{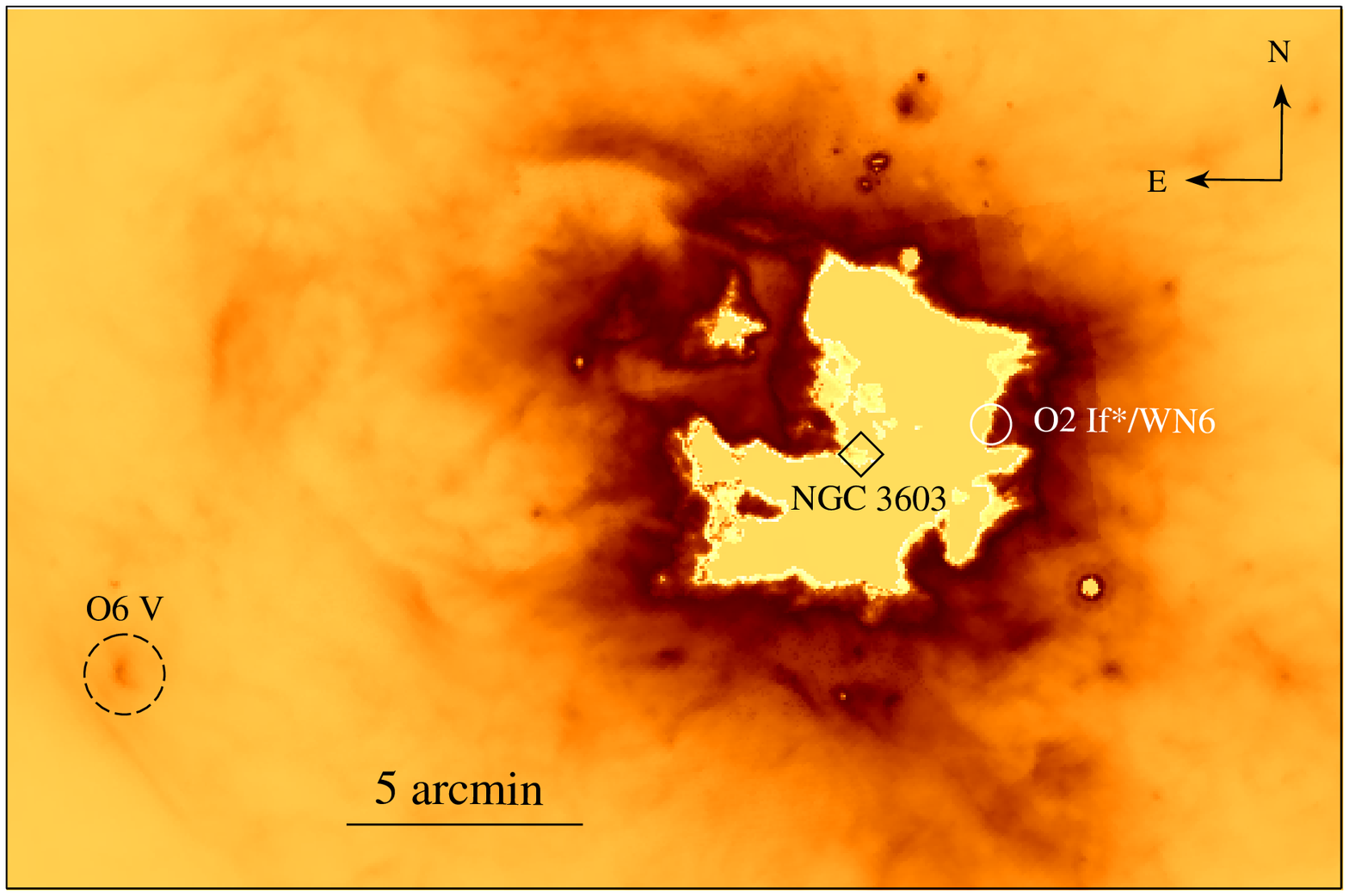}} \\
\end{minipage}
\caption{{\it Left}: MIPS 24\,$\mu$m image of the bow shock
detected near NGC\,3603. The position of the associated O6\,V star
(2MASS\,J11171292$-$6120085) is marked by a circle. {\it Right}:
MIPS $24\,\mu$m image of the highly saturated \hii region
associated with the star cluster NGC\,3603 (indicated by a
diamond). The positions of the bow-shock-producing star and the
O2\,If*/WN6 star (WR42e) are marked by circles (see text for
details). At the distance of NGC\,3603 of 7.6 kpc, 30 arcsec and 5
arcmin correspond to $\approx$1.1 and 10.9 pc, respectively.}
\label{fig:NGC3603}
\end{figure*}

Young massive star clusters lose their massive stellar content at
the very beginning of their evolution because of dynamical
few-body interactions (Gvaramadze, Kroupa \& Pflamm-Altenburg
2010b; Fujii \& Portegies Zwart 2011; Banerjee, Kroupa \& Oh
2012). The ejected stars form the population of field OB stars
(e.g. Gies 1987), whose space velocities range from $\sim$$10 \,
\kms$ (the escape velocity from the potential well of the parent
cluster) to several hundreds of $\kms$ (e.g. Heber et al. 2008).
About 20 per cent of high-velocity ($>$$30 \, \kms$) field OB
stars (the so-called runaway stars; Blaauw 1961) are moving
supersonically through the ambient interstellar medium (e.g.
Huthoff \& Kaper 2002) and produce bow shocks, which can be
detected in the optical (Gull \& Sofia 1979), infrared (van Buren
\& McCray 1988), radio (Benaglia et al. 2010), and X-ray
(L\'opez-Santiago et al. 2012) wavebands.

Detection of bow shocks around massive star clusters allows us to
reveal OB stars running away from these clusters (Gvaramadze \&
Bomans 2008; Gvaramadze et al. 2011c). Follow-up spectroscopy of
stars selected in this way and the geometry of their bow shocks
provide a powerful tool for linking the newly identified massive
stars to their parent clusters even in those cases when the proper
motion measurements for the stars are unavailable or unreliable
(Gvaramadze \& Bomans 2008; Gvaramadze et al. 2010a,b, 2011c;
Gvaramadze, Pflamm-Altenburg \& Kroupa 2011a). This in turn
provides useful constraints on the modelling of dynamical
evolution of young star clusters (e.g. Kroupa 2008; Portegies
Zwart, McMillan \& Gieles 2010) and has important consequences for
understanding the origin of the field OB stars (Gvaramadze et al.
2010b, 2012a). Bow shocks can also be used to constrain mass-loss
rates of their associated stars (Gull \& Sofia 1979; Kobulnicky,
Gilbert \& Kiminki 2010; Gvaramadze, Langer \& Mackey 2012b) and
could serve as probes of the Galactic magnetic field (Gvaramadze
et al. 2011b,c).

In this Letter, we report the discovery of a bow-shock-producing
star in the vicinity of the star cluster NGC\,3603
(Section\,\ref{sec:sea}). Follow-up optical spectroscopy of this
star led to its classification as O6\,V (Section\,\ref{sec:spec}).
In Section\,\ref{sec:dis}, we show that the relative position of
the O6\,V star and a recently discovered O2\,If*/WN6 star on the
sky is consistent with the possibility that both objects were
ejected from NGC\,3603 via the same dynamical event -- a
three-body encounter.

\begin{figure*}
\includegraphics[width=8cm,angle=270]{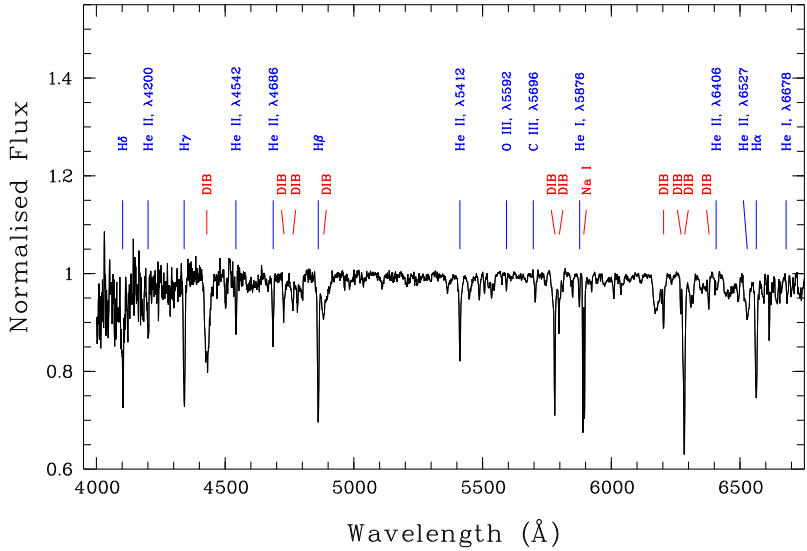}
\caption{Normalized spectrum of 2MASS\,J11171292$-$6120085 with
principal lines and most prominent DIBs indicated.}
\label{fig:spec}
\end{figure*}

\section{Search for bow shocks around NGC\,3603}
\label{sec:sea}

\subsection{NGC\,3603}
\label{sec:ngc}

NGC\,3603 is a very young ($\sim$2 Myr; Kudryavtseva et al. 2012)
and one of the most massive ($\sim$10$^4 \, \msun$; Harayama,
Eisenhauer \& Martins 2008) star clusters known in the Milky Way.
It is located in the Carina arm at a distance of $d$=7.6 kpc
(Melena et al. 2008). The cluster contains numerous O-type stars
and three very massive WN-type stars (Moffat, Drissen \& Shara
1994; Schnurr et al. 2008), whose initial masses could be as large
as $\approx$140$-170 \, \msun$ (Crowther et al. 2010). All three
WN-type stars are located in the cluster's core and two of them
are short-period ($\approx$4$-9$ day) binary systems (Schnurr et
al. 2008).

The existence of very massive (binary) stars in the cluster and
compactness of its core (the radius of the core is $\approx$0.2
pc; Harayama et al. 2008) suggest that, at least in the recent
past, NGC\,3603 was effective in producing massive runaway stars
(e.g. Gvaramadze et al. 2009; Gvaramadze \& Gualandris 2011). This
makes attractive a search for OB stars running away from this
cluster. However, given the large distance to NGC\,3603, it is
unlikely that the existing astrometric catalogues would provide
reliable proper motion measurements for stars ejected from the
cluster, unless their peculiar velocities exceed several hundreds
of $\kms$. The runaway status of stars around NGC\,3603 can, in
principle, be revealed through detection of their high peculiar
radial velocities, but in the absence of reliable proper motion
measurements for these stars it would be impossible to
unambiguously prove their relationship to the cluster. This makes
bow shocks the most important signature of massive stars running
away from NGC\,3603.

\subsection{Bow-shock-producing star 2MASS\,J11171292$-$6120085}
\label{sec:bow}

To search for bow shocks around NGC\,3603, we used archival {\it
Spitzer Space Telescope} 24\,$\mu$m image of NGC\,3603 and its
surroundings (Program Id.: 41024, PI: L.~Townsley) obtained with
the Multiband Imaging Photometer for {\it Spitzer} (MIPS; Rieke et
al. 2004). This image represents an $\approx 0\fdg5 \times 2\fdg0$
strip extended almost in the eastwest direction and centred at
$\approx$20 arcmin to the west of NGC\,3603. Visual inspection of
the image led to the discovery of five clear arc-like structures,
which we interpret as bow shocks (cf. Gvaramadze \& Bomans 2008;
Gvaramadze et al. 2011c). However, only one of them is opened
towards NGC\,3603 (see Fig.\,\ref{fig:NGC3603}) and therefore
could be generated by a massive star expelled from this cluster.
The stand-off distance of this bow shock is $\approx$7 arcsec (or
$\approx$0.25 pc).

Subsequent examination of the Digitized Sky Survey II (DSS-II;
McLean et al. 2000) red band image showed that the bow shock is
apparently generated by a star with coordinates: $\alpha _{\rm
J2000}$=$11^{\rm h} 17^{\rm m} 12\fs93, \delta _{\rm
J2000}$=$-61\degr 20\arcmin 08\farcs6$, which is located at
$\approx$0.262 degree (or $\approx$34 pc in projection) from the
centre of NGC\,3603. This star has a visual magnitude of
$\approx$15$-16$ (Zacharias et al. 2004; Lasker et al. 2008) and
2MASS (Two-Micron All Sky Survey) $J, H$, and $K_{\rm s}$
magnitudes of 11.79$\pm$0.03, 11.22$\pm$0.03, and 10.92$\pm$0.02,
respectively (Cutri et al. 2003). In what follows, we will use for
this star its 2MASS name -- 2MASS\,J11171292$-$6120085, or
J1117$-$6120, in short.

\section{Spectral type of J1117$-$6120}
\label{sec:spec}

\subsection{Spectroscopic observations and data reduction}
\label{sec:obs}

To determine the spectral type of J1117$-$6120 and thereby to
constrain its distance, we observed this star within the framework
of our programme of spectroscopic follow-up of candidate massive
stars revealed via detection of their bow shocks (e.g. Gvaramadze
et al. 2011c). For observation we used the Poor Weather time at
Gemini-South under the program-ID GS-2011A-Q-88. The spectroscopic
follow-up was performed with the GMOS-S (Gemini Multi-Object
Spectrograph South) in a long-slit mode, which provides coverage
from 3800 to 6750 \AA \, with a resolving power of $\sim$3000. The
spectrum of J1117$-$6120 was collected on 2011 February 20 under
fairly good conditions: with only some clouds and a $\approx$1
arcsec seeing. The aimed signal-to-noise ratio (S/N) of
$\approx$150 was achieved with a total exposure time of
3$\times$900 seconds.

The bias subtraction, flat-fielding, wavelength calibration and
sky subtraction were executed with the \textsf{gmos} package in
the \textsf{gemini} library of the {\sc iraf}\footnote{{\sc iraf}:
the Image Reduction and Analysis Facility is distributed by the
National Optical Astronomy Observatory (NOAO), which is operated
by the Association of Universities for Research in Astronomy, Inc.
(AURA) under cooperative agreement with the National Science
Foundation (NSF).} software. In order to fill the gaps in GMOS-S's
CCD, the observation was divided into three exposures obtained
with a different central wavelength, i.e. with a 5 \AA \, shift
between each exposure. The extracted spectrum was obtained by
averaging the individual exposures, using a sigma clipping
algorithm to eliminate the effects of cosmic rays. The average
wavelength resolution is $\approx$0.46 \AA~pixel$^{-1}$ (FWHM
$\approx$4.09 \AA), and the accuracy of the wavelength calibration
estimated by measuring the wavelength of ten lamp emission lines
is 0.061\,\AA. A spectrum of the white dwarf LTT\,3218 was used
for flux calibration and removing the instrument response.
Unfortunately, due to the weather conditions, any absolute
measurement of the flux is not possible.

\subsection{Spectral classification of J1117$-$6120}
\label{sec:clas}

Fig.\,\ref{fig:spec} presents the normalized spectrum of
J1117$-$6120 in the $\lambda\lambda 4000-6750$ \AA \, region,
where most of the identified lines are marked. The spectrum is
dominated by absorption lines of H\,{\sc i} and He\,{\sc ii},
which is typical of O-type stars. He\,{\sc i} $\lambda\lambda5876,
6678$, O\,{\sc iii} $\lambda 5592$ and C\,{\sc iii} $\lambda 5696$
absorptions are also clearly visible in the spectrum. The
relatively high interstellar extinction towards the star ($\approx
6$ mag; see Section\,\ref{sec:bir}) is manifested in numerous
diffuse interstellar bands (DIBs). A blend of strong Na\,{\sc i}
$\lambda 5890, 96$ absorption lines is of interstellar origin as
well.

The poor observational conditions and the extinction towards
J1117$-$6120 degraded the blue part of the spectrum, which
precludes us from using traditional classification criteria.
Instead, we utilized the classification scheme for dwarf O stars
in the yellow-green ($\lambda\lambda 4800-5420$ \AA) proposed by
Kerton, Ballantyne \& Martin (1999). Using their equation (3) and
the observed equivalent width (EW) of the He\,{\sc ii}
$\lambda$5412 line of 0.99$\pm$0.06 \AA, we found that the star is
of $\approx$O6\,V type, although the uncertainty of the
calibration leaves the possibility that it is either of 05\,V or
O6.5\,V type. Alternatively, the spectral type of J1117$-$6120 can
be derived using the EW(H$\gamma$)$-$absolute magnitude
calibration by Balona \& Crampton (1974), which for
EW(H$\gamma$)=2.28$\pm$0.06 \AA \, gives the same spectral type of
O6\,V.

\section{Discussion}
\label{sec:dis}

\subsection{Parent cluster of J1117$-$6120}
\label{sec:bir}

To estimate the distance to J1117$-$6120, one can use the observed
photometry of this star and the synthetic photometry of Galactic O
stars by Martins \& Plez (2006). The existing catalogues
(available through the VizieR catalogue access tool) provide quite
different measurements of optical magnitudes, e.g. $B$=17.3 mag
(USNO-A2.0; Monet 1998), $V$=15.7 mag (GSC2.3; Lasker et al.
2008), and $B$=16.2 mag and $V$=15.0 mag (NOMAD; Zacharias et al.
2004). In the lack of quality optical photometry, we will use only
the 2MASS photometry (see Section\,\ref{sec:bow}), whose $J$ and
$K_{\rm s}$ magnitudes are consistent within the margins of error
with those from the DENIS (Deep Near Infrared Survey of the
Southern Sky) database (DENIS Consortium 2005).

Using the absolute $K_{\rm s}$-band magnitude and the intrinsic
$J$$-$$K_{\rm s}$ colour typical of O6\,V stars of $-$4.13 mag and
$-$0.21 mag, respectively, we calculated the $K_{\rm s}$-band
extinction of $A_{K_{\rm s}}$=0.71$\pm$0.02 mag and the distance
to the star of 7.4$_{-0.5} ^{+0.9}$ kpc\footnote{Note the
dominating source of error in the distance is the uncertainty in
the spectral classification.}, which agrees well with the distance
to NGC\,3603 of 7.6 kpc. The extinction towards J1117$-$6120 can
also be estimated by matching the dereddened spectral slope of
this star with those of stars of similar effective temperature. In
doing so, we found the colour excess $E(B-V)$=2.07$\pm$0.05 mag.
Then using the extinction law from Rieke \& Lebofsky (1985) and
the standard total-to-selective absorption ratio $R_V$=3.1, we
found $A_V$=6.42$\pm$0.16 mag and $A_{K_{\rm s}}$=0.72$\pm$0.02
mag. The latter estimate is in excellent agreement with that based
on the 2MASS photometry.

Alternatively, the extinction can be estimated by using the
correlation between the intensity of the DIBs and $E(B-V)$ (see
Herbig 1995 for a review). Using EWs of DIBs at $\lambda 5780$ and
$\lambda 5797$ of 1.46$\pm$0.04 and 0.44$\pm$0.04 \AA,
respectively, and the relationships given in Herbig (1993), we
calculated $E(B-V)=2.78$$\pm$0.32 and 3.08$\pm$0.43 mag. These
estimates somewhat differ from those based on the spectral slope
and 2MASS photometry. This discrepancy could be caused by a
foreground region of enhanced number density of carriers of the
DIBs (cf. Gvaramadze et al. 2012c).

These estimates along with the orientation of the bow shock
strongly suggest that NGC\,3603 is the parent cluster of
J1117$-$6120. Unfortunately, this suggestion cannot be proved by
proper motion measurements because of the large distance to
J1117$-$6120. Indeed, inspection of the VizieR database showed
that none of the existing astrometric catalogues provide
significant proper motion measurements for this star.

\subsection{J1117$-$6120: a runaway star from a three-body encounter?}
\label{sec:run}

The young age of NGC\,3603 implies that J1117$-$6120 was ejected
dynamically, either because of a binary-binary or binary-single
encounter in the cluster's core. In the first case, the most
common outcome of the encounter is the exchange of the more
massive components into a new eccentric binary and ejection of the
less massive ones with high velocities (in general, the
trajectories of the ejected stars make an arbitrary angle with
each other). In the second case, a single star (usually the lowest
mass star among the stars participating in the encounter) is
ejected with a high velocity, while the binary system recoils in
the opposite direction to the single star. In both cases, the
ejected (high-velocity) stars gain their kinetic energy at the
expense of the increased binding energy of the post-encounter
binary, which ultimately could merge into a single star if its
orbit is sufficiently compact. Thus, if J1117$-$6120 was ejected
in the field via a three-body encounter then a massive binary or a
single merged star should exist on the opposite side of NGC\,3603
(cf. Gvaramadze \& Gualandris 2011).

Interestingly, such a star does indeed exist. This O2\,If*/WN6
star, called WR42e, was recently discovered by Roman-Lopes (2012).
It is located at $\theta_1$$\approx$0.045 degree west-northwest of
NGC\,3603 (see Fig.\,\ref{fig:NGC3603}), just on the opposite side
of J1117$-$6120 (recall that this star is separated from NGC\,3603
by $\theta _2$$\approx$0.262 degree). If both stars were ejected
from NGC\,3603 via a three-body encounter, then WR42e should be a
very massive binary or a single merged star. Moreover, one can
``weigh" this star using the conservation of the linear momentum.
Assuming the mass of the O6\,V star of $m$$\approx$$30 \, \msun$
(e.g. Martins, Schaerer \& Hillier 2005), one has that at the time
of encounter the mass of the recoiled binary (now the O2\,If*/WN6
star) was $M$=$(\theta_2 /\theta_1 )m$$\approx$$175 \, \msun$.

Although the existing proper motion measurements for both stars
are very unreliable\footnote{Note that the huge proper motion of
WR42e given in the PPMXL catalogue (R\"{o}ser, Demleitner \&
Schilbach 2010) is erroneous (S.R\"{o}ser, personal
communication).}, one can constrain their peculiar velocities
using the following arguments. The minimum space velocity of a
star leaving its parent cluster should exceed the escape velocity,
$v_{\rm esc}$, from the cluster's potential well, which for
NGC\,3603 is $\approx$$10 \, \kms$ (here we assumed that at the
time of the three-body encounter, say 1 Myr ago, the radius of the
cluster was $\sim$1 pc). Correspondingly, the recoil velocity of
WR\,42e is $v_1$$\geq$$v_{\rm esc}$, while the ejection velocity
of J1117$-$6120 is $v_2$=$(\theta_2 /\theta_1)v_1$$\geq$$60 \,
\kms$. These estimates have important consequences for
understanding the nature of WR42e (see next section).

The velocity estimate for J1117$-$6120 can also be used to
constrain the number density of the ambient interstellar medium,
$n_0$. Using the wind mass-loss rate and terminal velocity typical
of O6\,V stars ($\dot{M}=10^{-7} \msun \, {\rm yr}^{-1}, v_\infty
=2500 \, \kms$; Mokiem et al. 2007), one finds $n_0 \leq 2.6 \,
{\rm cm}^{-3}$.

\subsection{WR42e as a merged binary star}

The ejection velocity of J1117$-$6120 should be compared with a
typical ejection velocity resulting from encounters between a very
massive binary and a single (less massive) star, which for
binaries with equal-mass components is $\approx$$0.8v_{\rm orb}$
(Hills \& Fullerton 1980), where $v_{\rm orb}$ is the orbital
velocity in the binary. This comparison implies that to expel
J1117$-$6120 with the velocity of $\geq$$60 \, \kms$, the binary
separation should be $\leq$15 AU.

A massive binary of this small separation would be a source of
strong X-ray emission (e.g. Usov 1992). Following Crowther et al.
(2010; see their Section\,5.1 and references therein), one can
estimate the expected X-ray luminosity, $L_{\rm X} ^{\rm exp}$, of
shocked winds in WR42e assuming that this system is composed of
two equal-mass components (with the mass-loss rates of
$\sim$$2\times10^{-5} \, \msun \, {\rm yr}^{-1}$ and wind
velocities of $2600 \, \kms$) and that 10 per cent of the shock
energy contributes to the X-ray emission. We found $L_{\rm X}
^{\rm exp}$$\approx$$5.6\times 10^{34} \, {\rm erg} \, {\rm
s}^{-1}$, which is more than two orders of magnitude larger than
the observed X-ray luminosity of WR42e of $L_{\rm X} ^{\rm
obs}$$\approx$$2.3$$\times$$10^{32} \, {\rm erg} \, {\rm s}^{-1}$
(Roman-Lopes 2012). Moreover, $L_{\rm X} ^{\rm obs}$ comprises a
fraction of $\sim$2$-$4$\times$$10^{-8}$ of the bolometric
luminosity of the star (see below), which is typical of single
stars (Chlebowski, Harnden \& Sciortino 1989). From this it
follows that WR42e might be a merger product of the recoiled
binary system, which is coalesced because of the encounter
hardening.

Numerical simulations by Suzuki et al. (2007) showed that during
the merger process the system loses about 10 per cent of its mass,
which in the case of WR42e corresponds to $\sim$$20 \, \msun$. The
resulting very massive star is a fast rotator with
higher-than-average helium abundance, which might be responsible
for the Of*/WN-type spectrum of WR42e (cf. Walborn et al. 2010).
During the subsequent $\sim$1 Myr the star additionally loses
about 20$-$$30 \, \msun$ in the form of stellar wind, so that its
current mass should be $\approx$$125$$-$$135 \, \msun$, which
corresponds to a luminosity of
$\log(L/L_{\odot})$$\approx$$6.3$$-$$6.5$ (Crowther et al. 2010;
Ekstr\"{o}m et al. 2012). This value should be compared with the
luminosity of WR42e. Using the 2MASS $J$ and $K_{\rm s}$
magnitudes of this star of 10.18 and 9.04, respectively, and
($J$$-$$K_{\rm s}$)$_0$=$-0.21$ mag (Martins \& Plez 2006), one
finds $M_{K_{\rm s}}$$=$$-6.25$ mag, which for the $K$-band
bolometric correction of $-$(4.4$\div$5.2) mag (Crowther \&
Walborn 2011) corresponds to $\log(L/L_{\odot})$$\approx$6.2$-$6.5
(cf. Roman-Lopes (2012).

To conclude, if our proposal on the relationship between
J1117$-$6120 and WR42e is correct then one can expect that the
peculiar radial velocity of the latter star should be about six
times smaller than that of the former one. For J1117$-$6120 we
measured a heliocentric radial velocity of 21.4$\pm$$6.3 \, \kms$,
which is an average over the hydrogen and He\,{\sc ii} lines.
After correction for the Galactic differential rotation and the
solar peculiar motion, we found the peculiar radial velocity of
J1117$-$6120 of $-4.8 \, \kms$, i.e. the star is moving almost in
the plane of sky. [This estimate was derived using the Galactic
constants $R_0$=8.0 kpc and $\Theta _0$=$240 \, \kms$ (Reid et al.
2009) and the solar peculiar motion
$(U_{\odot},V_{\odot},W_{\odot})=(11.1,12.2,7.3) \, \kms$
(Sch\"onrich, Binney \& Dehnen 2010).] Correspondingly, the
peculiar radial velocity of WR42e should also be almost zero.
Radial velocity measurements for WR42e are therefore of crucial
importance for testing our proposal.

\section{Acknowledgements}

We are grateful to L. Kaper (the referee) for his comments on the
manuscript. AYK acknowledges support from the National Research
Foundation of South Africa. This work has made use of the
NASA/IPAC Infrared Science Archive, which is operated by the Jet
Propulsion Laboratory, California Institute of Technology, under
contract with the National Aeronautics and Space Administration,
the SIMBAD database and the VizieR catalogue access tool, both
operated at CDS, Strasbourg, France.

\end{document}